\documentclass[12pt,preprint]{aastex}           %% 1-column, 1-spaced
   \newcommand{\tnm}{\tablenotemark}
   \newcommand{\gsim}{\rlap{$>$}{\lower 1.0ex\hbox{$\sim$}}}

   \renewcommand{\sec}{\prime\prime}
   \renewcommand{\min}{\prime}
   \renewcommand{\deg}[0]{\circ}

   \shorttitle{Spitzer 24micron Observations of EROs}
   \shortauthors{Yan et al}

   \normalsize

\begin{document}

%\title{24~micron Mid-infrared Properties of Optical/near-IR Selected 
%Extremely Red Galaxies}
%\title{Assembly of Massive Galaxies at $z \sim 1 - 2$: 
%Evidence from the Spitzer 24micron Observations
%in the First Look Survey ELAIS N1 Field} 

\title{Spitzer 24micron Observations of Optical/Near-IR Selected
Extremely Red Galaxies: Evidence for Assembly of Massive Galaxies at $z \sim 1 - 2$?}

\author{Lin Yan\altaffilmark{1}, Philip I. Choi\altaffilmark{1}, D. Fadda\altaffilmark{1}, 
F.R. Marleau\altaffilmark{1}, B.T. Soifer\altaffilmark{1,2} \\
M. Im\altaffilmark{3}, L. Armus\altaffilmark{1}, D.T. Frayer\altaffilmark{1}, L.J. Storrie-Lombardi\altaffilmark{1}, 
D.J. Thompson\altaffilmark{2},
H.I. Teplitz\altaffilmark{1}, G. Helou\altaffilmark{1}, P.N. Appleton\altaffilmark{1},
S. Chapman\altaffilmark{2}, F. Fan\altaffilmark{1}, I. Heinrichsen\altaffilmark{1}, M. Lacy\altaffilmark{1}, D.L. Shupe\altaffilmark{1}, G.K. Squires\altaffilmark{1}, J. Surace, G. Wilson\altaffilmark{1}}

\altaffiltext{1}{Spitzer Space Telescope Science Center, California
  Institute of Technology, 1200 East California Boulevard, MS 220-6,
  Pasadena, CA 91125; Send offprint requests to Lin Yan:
lyan@ipac.caltech.edu.}

\altaffiltext{2}{The Caltech Optical Observatories, Caltech, Pasadena,  CA~91125}

\altaffiltext{3}{School of Earth and Environmental Sciences, Seoul National University, Shillim-dong, Kwanak-gu,Seoul, S. Korea}

\begin {abstract}
  
  We carried out the direct measurement of the fraction of dusty sources in a 
  sample of extremely red galaxies with ($R - K_s) \ge 5.3$mag and 
  $K_s < 20.2$mag, using 24$\mu m$ data from the 
  {\it Spitzer Space Telescope}. Combining deep 24$\mu m$, $K_s$- and
  $R$-band data over an area of $\sim$64~sq.arcmin in ELAIS N1 of
  the Spitzer First Look Survey (FLS), 
  we find that 50$\pm6$\%\ of our ERO sample have measurable 24$\mu m$ flux
  above the 3$\sigma$ flux limit of 40$\mu$Jy. This flux limit
  corresponds to a SFR of 12$M_\odot$/yr at $z \sim 1$, much more
  sensitive than any previous long wavelength measurement.
  The 24$\mu m$-detected EROs have
  24-to-2.2 and 24-to-0.7$\mu m$ flux ratios consistent with
  infrared luminous, dusty sources at $z \ge 1$, and an 
  order of magnitude too red to be explained by an infrared quiescent spiral 
  or a pure old stellar population at any redshift. 
  Some of these 24$\mu m$-detected EROs
  could be AGN, however, the fraction among the whole ERO sample is probably
  small, 10-20\%, as suggested by deep X-ray observations
  as well as optical spectroscopy. Keck optical spectroscopy of a sample of similarly 
  selected EROs in the FLS field suggests that most of the 
  EROs in ELAIS N1 are probably at $z \sim 1$.  
  The mean 24$\mu m$ flux (167$\mu$Jy) of the 24$\mu m$-detected ERO sample 
  roughly corresponds to the rest-frame
  12$\mu m$ luminosity ($\nu L_{\nu}(12\mu m)$) of $3 \times 10^{10} L_\odot$ at $z \sim 1$.
  Using the correlation between IRAS $\nu L_{\nu}(12\mu m)$ and
  infrared luminosity $L_{IR}(8-1000\mu)$, we infer that the
  $<L_{IR}>$ of the 24$\mu m$-detected EROs is $3 \times 10^{11}
  L_\odot$ and $10^{12} L_\odot$ at $z=1.0, 1.5$ respectively, similar to
  that of local LIRGs and ULIGs. The corresponding SFR would be
  roughly $50 - 170 M_\odot$/yr. If the time scale of this starbursting phase
  is on the order of $10^8$yr as inferred for the local LIRGs and ULIGs, the
  lower limit on the masses of these 24$\mu m$-detected EROs 
  is $5\times 10^9 - 2\times 10^{10} M_\odot$. It is plausible that some of the
  starburst EROs are in the midst of violent transformation to become massive
  early type galaxies at the epoch of $z \sim 1 - 2$.

\end{abstract}

\keywords{galaxies: bulges --
          galaxies: starbursts --
          galaxies: infrared luminous --
          galaxies: galaxy evolution --
          galaxies: high-redshifts}

\section{Introduction}

Optical/near-IR colors, such as $(R - K_s)$ or $(I - K_s)$ have been
commonly used in wide-area surveys to select old stellar populations
at $z \sim 1-2$ (Cimatti et al.\ 2002; McCarthy et al.\ 2001; McCarthy
2004 for the review of this subject).  Near-IR observations, which
sample cool low-mass stars, are sensitive to old stellar populations.
The ($R - K_s)$ of 5.3 corresponds to the calculated color
of a passively evolving elliptical galaxy at $z = 1$. Therefore, in principle, 
the color criterion of ($R - K_s) \ge 5.3$mag or ($I
- K_s) \ge 4$mag (EROs) should select early type galaxies at $z \sim 1$.
However, these color
selections are also sensitive to dust-reddened, star-forming systems, 
and examples of both passively evolving ellipticals and dusty starburst
EROs have been found (Soifer et al.\  1999; Hu \&\ Ridgway 1994;
McCarthy, Persson \&\ West 1992; Graham \&\ Dey 1996).  This indicates
that the optical/near-IR SEDs of these sources are sufficiently
degenerate that these color criteria cannot effectively distinguish
between them. In addition, a small fraction of EROs (10-20\%) could also be AGN, 
as shown by deep Chandra data and optical spectroscopy 
(Alexander et al. 2002; Yan, Thompson \&\ Soifer 2004).
The relative contribution of these two galaxy types ---
old stellar populations and dust-reddened, star-forming galaxies ---
is a critical issue for many surveys whose goal is to determine the
evolution of the mass function and the formation of massive galaxies.
Deep optical spectroscopy indicates that a large fraction (30-50\%) of
EROs have emission lines (Cimatti et al.\ 2002; Yan, Thompson \&\ 
Soifer 2004; McCarthy et al.\  2004); however, it remains unclear what
fraction of EROs are truly dust obscured galaxies.

MIPS 24$\mu m$ data from the {\it Spitzer Space Telescope} (Rieke et
al.  2004; Werner et al.\ 2004) offer the first opportunity to directly
address this critical issue.  Dusty, star-forming galaxies are clearly
distinguished from early-type galaxies at mid-IR wavelengths.  Between
$z \sim 1-2$, the MIPS data is especially discriminating as strong,
rest-frame 6$-$12$\mu$m polycyclic aromatic hydrocarbon (PAH) dust
features redshift into the 24$\mu m$-band.  In this Letter, we present
our initial study of the 24$\mu m$ properties of the ($R - K_s \ge
5.3$)mag selected EROs in ELAIS N1.  Throughout the paper,
we adopt $H_0=70$km/s/Mpc, $\Omega_{M} = 0.3$, $\Omega_{\Lambda} =
0.7$, and the Vega system for optical/NIR magnitudes.

\section{Data}

\subsection{Spitzer 24$\mu m$ Observations and Data Reduction}

The primary dataset used in this Letter is in the ELAIS N1 field,
which is a part of the Spitzer First Look Survey (FLS)\footnote{For
details of the FLS observation plan and the data release, see
http://ssc.spitzer.caltech.edu/fls.}.  The field was observed in a
$2\times2$ photometry mode mosaic, and the raw data were processed and
stacked by the data processing pipeline at the Spitzer Science Center
(SSC).  Source catalogs at 24$\mu m$ were generated using StarFinder
(Diolaiti et al.\ 2000), which measures profile-fitted fluxes for
point sources.  The complete description of the 24$\mu m$ data
reduction and source catalog can be found in Fadda et al.\ (2004b) and
Marleau et al. (2004).  To aid in the interpretation of the results
from ELAIS N1, we also analyzed the FLS verification strip (FLSV).
The data presented here covers 64~sq.arcmin in ELAIS N1 and
256~sq.arcmin in the FLSV.  Table 1 summarizes the salient
characteristics of all data used in this paper.

\subsection{Optical, Near-IR Imaging and Keck Spectroscopy}

All $R$-band observations were taken at the Kitt Peak National 
Observatory.  Final stacked images and source catalogs have been 
publicly released (Fadda et al. 2004a). The $K_s$-band data were
obtained using the Wide-Field Infrared Camera (WIRC) on the Palomar
200-inch telescope. The $K-band$ data in Elais N1 covers $8^{\min}\times8^{\min}$.
A detailed description of the WIRC observations and data reduction will
included in a separate paper by Choi et al.\ (2004).  Finally,
high-resolution, optical spectra of a $K_s$-selected galaxy sample
within the FLS region were obtained using the Keck, Deep Imaging
Multi-Object Spectrograph (DEIMOS) (Faber et al.\ 2003).  A total of
$\approx1000$ redshifts were measured, of which 112 (52 EROs) are
included in the analysis in this paper.

\section{Results and Implications}

\subsection{The 24$\mu m$ detected EROs} 

To merge the R/K and $24\mu m$ source catalogs, we used 
a simple positional matching method with a $2.4^{''}$ match radius,
which corresponds to $3\sigma$ combined astrometric uncertainty
from the $K_s$ and $24\mu m$ data. Due to the relatively low source
density (7arcmin$^{-2}$), the likelihood of spurious matches
is small (3\%), consistent with the fact
that we find only one multiple match out of 65. In addition, 
we test the robustness of each match by comparing
the probability of the measured separation based on the astrometric
uncertainties to the probability of a
spurious detection based on the 24$\mu m$ source density. The probability ratios
above unity imply to be likely real matches, and we found that all but two
matches meet this criteria. We retained these two matches since the visual
inspection suggests that they could still be associated with physical
offset centroids. Bright stars were rejected using $R$-band images. 
Stellar contamination in our sample is expected to be small since
the field is at the galactic latitude of 41$^\deg$, and
24$\mu m$ data samples the tail of the Rayleigh-Jeans energy distribution.

Since the $R$- and $K_s$-band data have similar seeing, ($R - K_s$)
colors were measured using a fixed $3^{\sec}$ diameter aperture.
The ERO catalog with ($R-K_s) \ge 5.3$mag and $K_s < 20.2$mag (6$\sigma$)
consists of 129 galaxies over 64~sq.arcmin in ELAIS N1. Figure 1
shows the $(R-K_s)$ vs. $K_s$ distribution for all sources ({\it black crosses}.
The ($R-K_s) \ge 5.3$mag and $K_s < 20.2$mag limits are shown
as {\it solid horizontal} and {\it solid vertical lines}, and sources with 24$\mu m$
detected counterparts are indicated as {\it red open circles}.
We find that 24$\mu m$ detected sources have slightly redder ($R - K_s$)
colors than non-detected sources, consistent with the expectation that
24$\mu m$ emission is an indicator of dust extinction.

Of the 129 EROs, 65 (50$\pm$6\%) have 24$\mu m$ emission with flux
greater than 40$\mu$Jy.  The fraction of 24$\mu m$ detected EROs
becomes slightly higher for redder sources, with 56$\pm15$\%\ and
67$\pm30$\%\ for ($R - K_s) \ge 6.0$mag and ($R - K_s) \ge 6.5$mag
respectively. However, the errors of these fractions are large due to 
the shallow $R$ band limit. Deeper data would be needed to reduce
the uncertainties. In figure 2a, we show the 24$\mu m$ flux 
distributions of both the total and ERO samples.  
We find that 70\%\ of the 24$\mu m$ EROs have
$f_{24\mu m} > 90\mu$Jy and that the mean 24$\mu m$ flux of the ERO
sample is 167$\mu$Jy. Figure 2b shows $K_s$ vs.  24$\mu m$ flux for all detected sources
({\it solid squares}) and for the EROs ({\it large open circles}).  We
find that amongst the 24$\mu m$ detected sources, $K_s$ is weakly
correlated with $f_{24\mu m}$, with large scatter. This is expected 
since these two bands sample light from different physical origins, 
stellar photosphere versus dust emission. This also explains why many faint
$K_s$ sources have fairly bright 24$\mu m$ fluxes. For $K_s$ bright ($\sim18 - 17$) 
and 24$\mu m$-faint sources, the mid-infrared data is deep enough to detect
the emission from normal galaxy populations at low redshifts.

\begin{figure}[h!]
\plotone{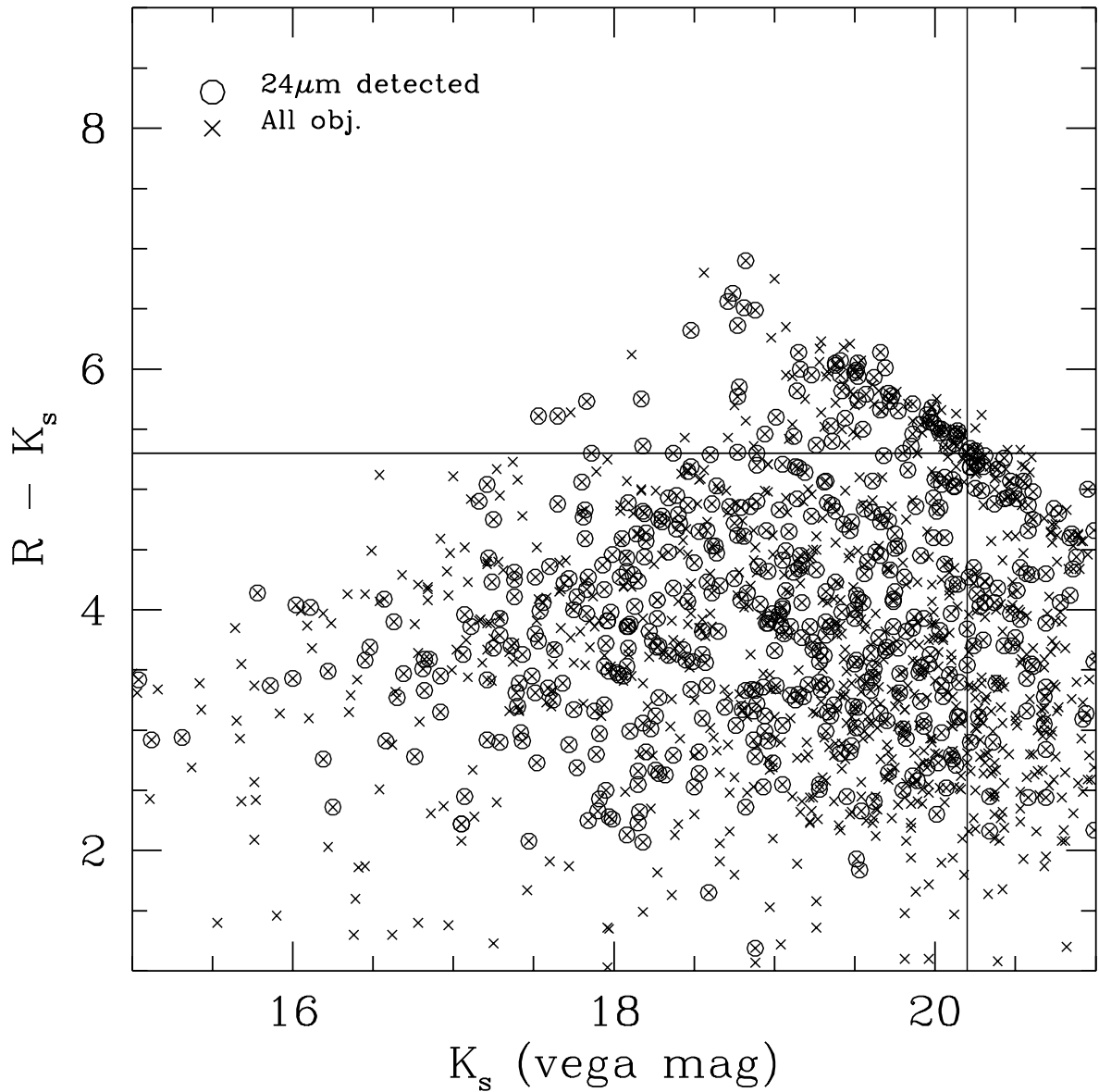}
\caption{The ($R - K_s$) vs. $K_s$ color-magnitude diagram. 
  All $K_s$-band detected sources are shown as {\it crosses},
  and sources with 24$\mu m$ counterparts are indicated by {\it 
    open circles}.  The ($R-K_s) \ge 5.3$mag and $K_s < 20.2$mag
  limits are shown as {\it solid horizontal} and {\it solid vertical}
  lines.  }
\end{figure}

\begin{figure}[h!]
\plotone{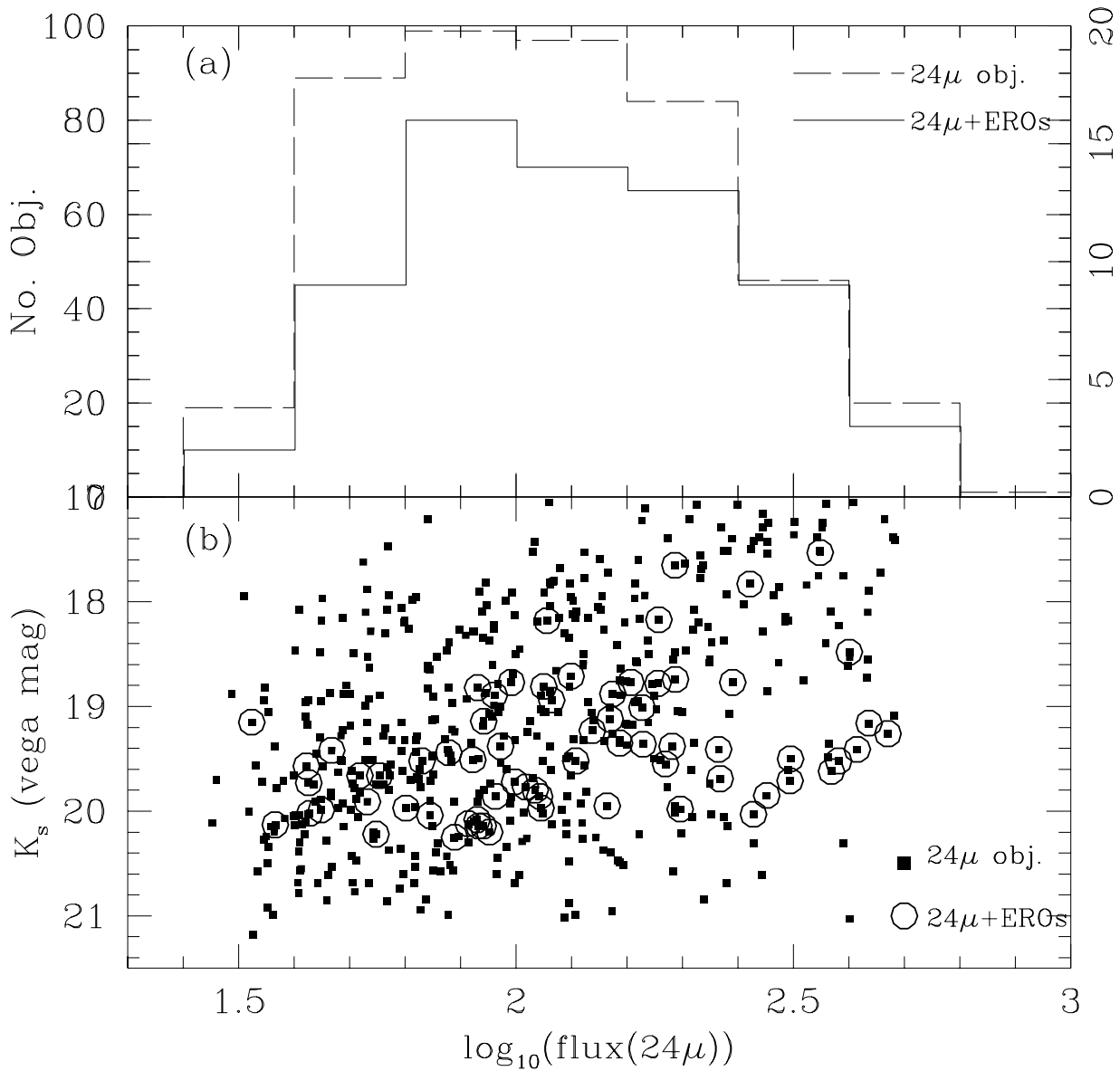}
\caption{{\bf Panel 2a} shows 
the 24$\mu m$ flux distribution of the total ({\it dashed line})
  and 24$\mu m$-detected ERO ({\it solid line}) populations.  {\bf Panel 2b} shows $K_s$
  vs. 24$\mu m$ flux for all detected sources ({\it solid squares})
  and for the EROs ({\it large open circles}).}
\end{figure} 

\subsection{The Nature of the EROs with 24$\mu m$ Emission}

What types of galaxies are the 24$\mu m$ detected EROs?  In Figure 3a,
we present a 24-to-0.7$\mu m$ and 24-to-2.2$\mu m$ color-color diagram
showing that 24$\mu m$-detected EROs ({\it red squares}) are clearly
separated from EROs without 24$\mu m$ emission ({\it green points}), as
well as the general non-ERO population ({\it black and blue}).  This
suggests that the MIR/Optical and MIR/NIR colors of 24$\mu m$-detected
EROs are unique, distinguishing them from other populations.  Figure
3b presents the expected colors as a function of redshift for various
SED templates, assuming no evolution.  The SEDs for M51 (normal
spiral), M82 (starburst) and Mrk231 (dusty AGN) are taken from Silva
et al.\ (1998), Fadda et al.\ (2002), and Chary \&\ Elbaz (2001).  In
the case of M31's bulge (old stellar population), we use the near-IR
J, H, K, IRAS 12, 25, 60 and 100$\mu$m fluxes, all within a 4$^{'}$
diameter aperture of the central nucleus (Soifer et al.\ 1986) and
merge this with a 10~Gyr theoretical SED from Bruzual \&\ Charlot.
\footnote{ftp://gemini.tuc.noao.edu/pub/charlot/bca5} 
For comparison,
a sample of 112, 24$\mu m$ sources from the FLSV with known
spectroscopic redshifts are marked to show the actual redshift range
of the 24$\mu m$ EROs in ELAIS N1. Of these 112 redshifts, 10 are
24$\mu m$-detected EROs with $ 0.8 < z_{spec} < 1.3$, and the optical
spectra of these 10 sources all have [OII]3727\AA emission line.

Comparing Figure 3a \& 3b, we reach the following conclusions: 
1) 24$\mu m$-detected EROs in ELAIS N1 with 
($R - K_s) \ge 5.3$mag, $K_s<20.2$mag, and $f_{\nu}(24\mu m$)$>40\mu$Jy 
are infrared bright sources at $z \ge 1$. They have colors similar to 
starbursts like M82 at $z \ge 1$, or dust reddened AGN like Mrk231 at $z \ge 0.7$.
Their colors are too red to be explained by any normal spiral or old stellar 
populations at any redshifts.
2) The likely redshift range of these 24$\mu m$ EROs is $z \ge 1$, as predicted
from the model SEDs of M82 and Mrk231. This is further confirmed
by comparison to the Keck spectroscopic sample from the FLSV.
3). The remaining half of the ERO population are probably galaxies with old stellar 
populations at $z \sim 1$, as suggested by the tracks in Figure 3b. 

\begin{figure}[h!]
\plotone{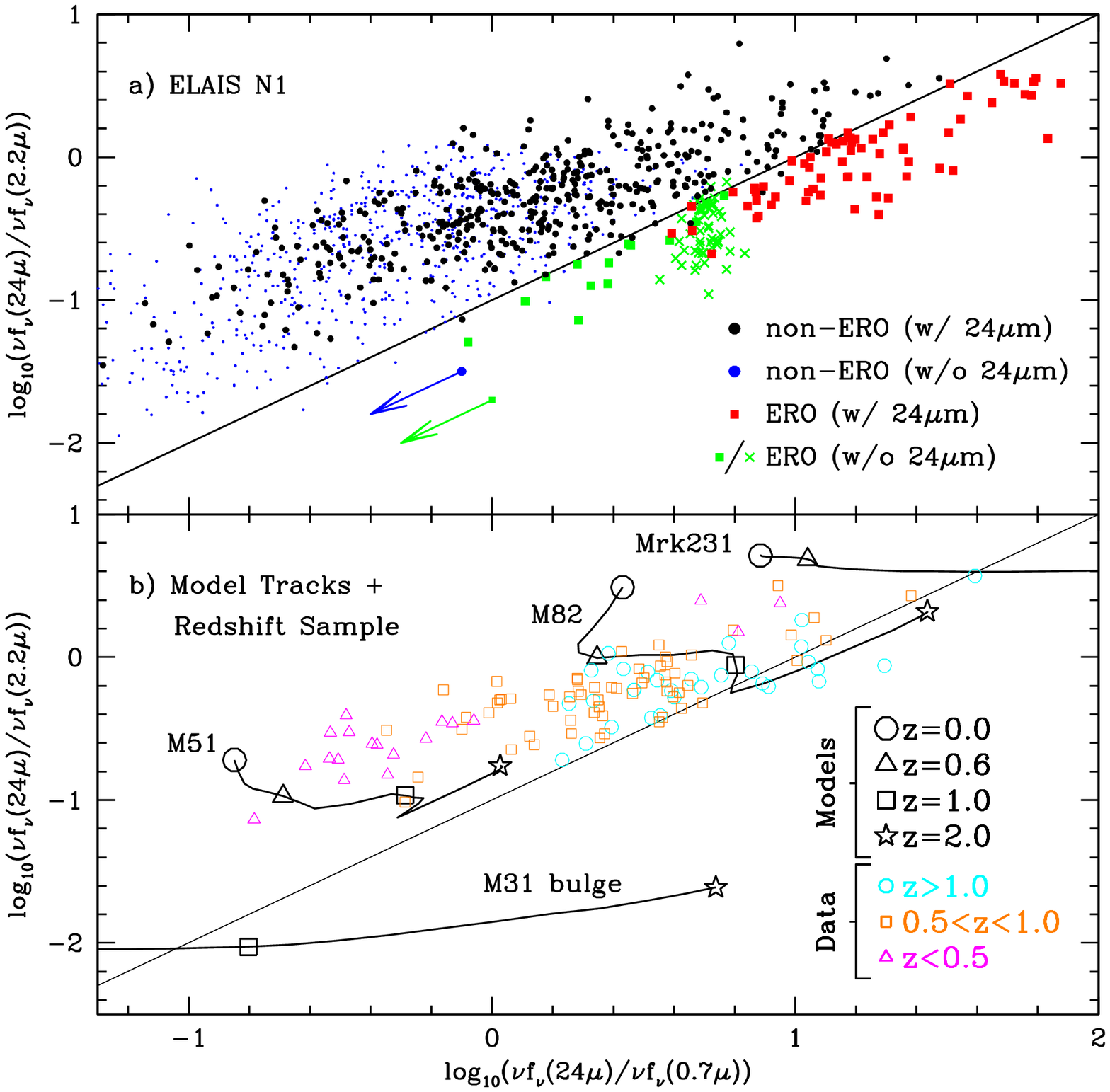}
\caption{The two panels show $\log_{10}(\nu f_{\nu}(24\mu m)/\nu f_{\nu}(0.7\mu m$)) 
  versus $\log_{10}(\nu f_{\nu}(24\mu m)/\nu f_{\nu}(8\mu m$)). {\bf Panel 3a}
  presents the data from ELAIS N1.  In this panel, {\it red/green
    squares} represent EROs with/without 24$\mu m$ detections.  For the
  non-24$\mu m$ EROs, the green crosses have only upper limits in the
  $R$-band.
  The {\it black/blue points} represent the full $K_s$-selected sample
  with/without 24$\mu m$ detections.  The plotted colors of sources not
  detected at 24$\mu m$ are upper limits and could move down to
  lower-right corner of the diagram, as illustrated by the {\it green
    \& blue arrows}.  The {\it black solid line} in both panels
  represents constant ($R - K_s) = 5.3$mag color.  {\bf Panel 3b}
  shows the computed color-color tracks as a function of redshift for
  a set of SED templates of local, well-known galaxies.  The subset of
  the FLSV sample with measured spectroscopic redshifts
  are overplotted for three different redshift bins: 
  $z < 0.5$ ({\it magenta open triangles}); $0.5 < z < 1.0$ 
  ({\it orange open squares}); and $z> 1.0$ ({\it cyan open circles}).
}
\end{figure}

Half of the ELAIS N1 ERO sample have 24$\mu m$ dust emission. 
The key question is what their infrared luminosities are.
In the previous section, we conclude that the 24$\mu m$-detected EROs
have colors similar to M82, however, Figure 3a does not set any constraints
on either their luminosities or masses.
We have a total of 112 redshifts for objects with 24$\mu m$
counterparts (the FLSV 3$\sigma$ flux limit is 90$\mu$Jy).  In Figure
4, we present the observed 24$\mu m$ luminosity versus redshift for the
spectroscopic FLSV sample. At $z \sim 1$ the observed 24$\mu m$
luminosity roughly corresponds to the rest-frame IRAS 12$\mu m$
luminosity, within a factor of 2. 
The Spitzer 24$\mu m$ filter is narrower than the IRAS 12$\mu m$ filter,
but ignoring this difference, a crude conversion of the observed
Spitzer 24$\mu m$ flux can be made to place a lower limit on the
rest-frame IRAS 12$\mu m$ luminosity.
We can infer the total infrared luminosity of 24$\mu m$ sources at $z
\sim 1$ by using the correlation between the 12$\mu m$ luminosity ($\nu
L_{\nu}(12\mu m)$) and the infrared luminosity $L_{IR}(8-1000\mu m)$,
$L_{IR} = 0.89\times (\nu L_{\nu}(12\mu m))^{1.094} L_\odot$ (Soifer et
al.\ 1989; Chary \&\ Elbaz 2001).  The mean flux (167$\mu$Jy)
of the 24$\mu m$-detected EROs in ELAIS N1 implies the $L_{IR} \sim 3 \times 10^{11}
L_\odot$ at $z=1.0$ and $L_{IR} \sim 10^{12} L_\odot$ at $z=1.5$, 
similar to that of local LIRGs and ULIGs.
The corresponding SFR is $50 - 170 M_\odot$/yr, using
$SFR = 1.71 \times 10^{-10} L_{IR}(M_\odot/yr)$ (Kennicutt 1998).
We emphasize that the 3$\sigma$ flux limit of 40$\mu$Jy in ELAIS N1
corresponds to a SFR of 12$M_\odot$/yr, more sensitive than the deepest
1.4GHz observation (Smail et al. 2002). 
In comparison, the SFR derived from
the rest-frame [OII]3727\AA emission line for EROs 
is roughly a few $\times M_\odot$/yr (Yan, Thompson \&\ Soifer 2004; Cimatti et al. 2003). 
Our results are consistent with very deep 1.4GHz (rms $\sim 3\mu$Jy) measurements
(Smail et al. 2002).

\begin{figure}[h!]
\plotone{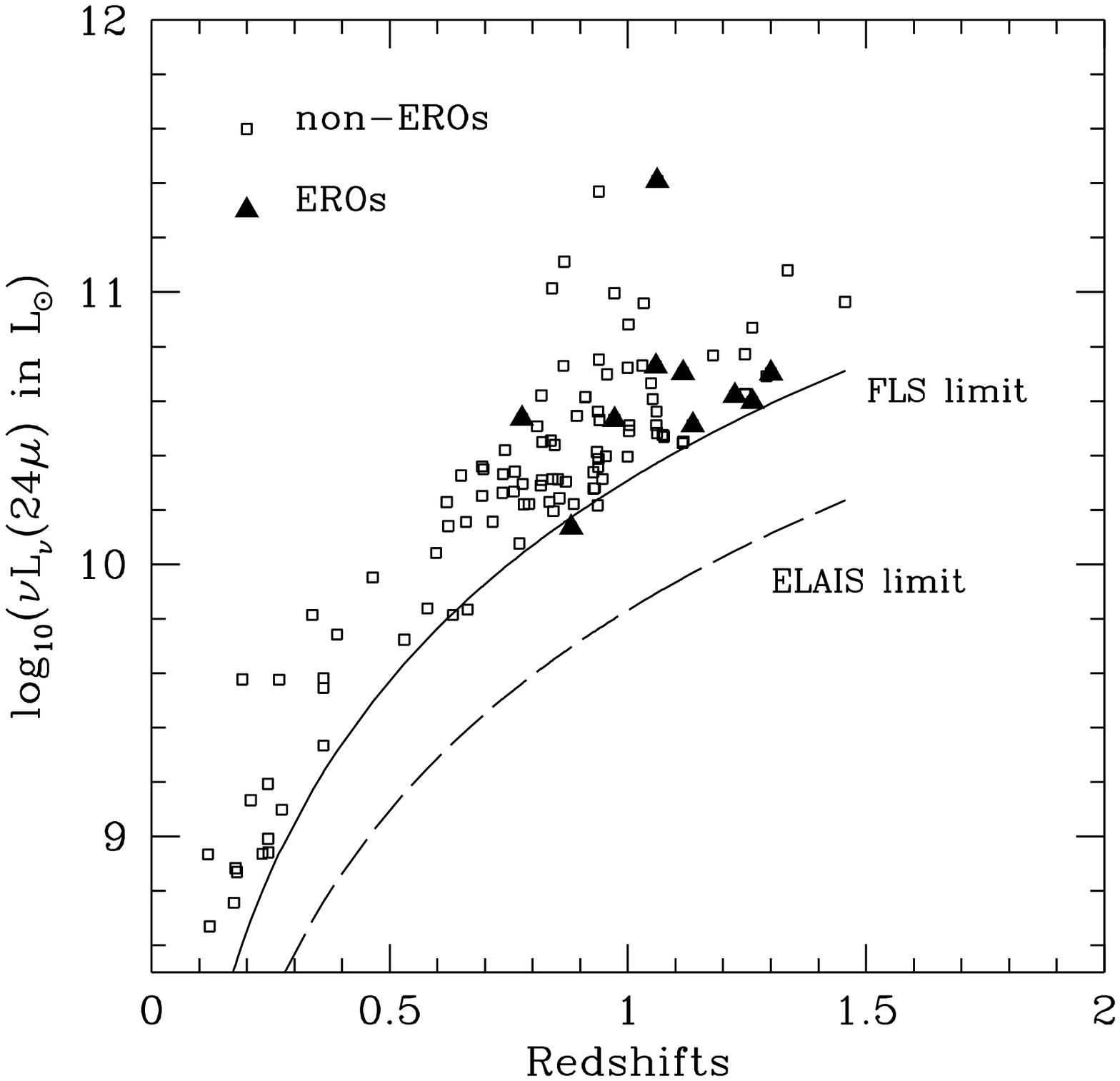}
\caption{The observed 24$\mu m$ luminosity as a function of spectroscopic redshift for
  a sample of 112 24$\mu m$-detected sources in the FLSV region. The
  solid triangles are 24$\mu m$-detected EROs.}
\end{figure}

\subsection{Summary \& Implications}

Spitzer 24$\mu m$ data in ELAIS N1 have revealed that 50$\pm6$\%\ 
of EROs with ($R - K_s) \ge 5.3$mag and $K_s <20.2$mag have detectable
24$\mu m$ emission above 40$\mu$Jy.
The colors and inferred redshifts of the 24$\mu m$-detected EROs
suggest that they are infrared-luminous, dusty sources at $z \ge 1$.
Their mean 24$\mu m$ flux (167$\mu$Jy) 
corresponds to $<L_{IR}> \sim 3\times 10^{11} L_\odot$ at $z \sim 1$
and to $<L_{IR}> \sim 10^{12} L_\odot$ at $z \sim 1.5$.  
Some of these 24$\mu m$-detected EROs could be AGN. 
The one mega-second Chandra observation in the HDF and
optical spectroscopy suggest that the fraction of EROs likely to be
AGN is small, around 10-20\%\ (Alexander et al. 2002; Yan, Thompson \&\ Soifer 2004).
These dusty AGN can be identified when we
combine this current analysis with the IRAC data.
Our result suggests that a significant fraction of EROs are
extremely active starbursts (LIRGs or ULIGs). If the time scale of
this starbursting phase is on the order of $10^8$/yr, as inferred from
local LIRGs and ULIGs (Sanders \&\ Mirabel 1996), we can set a lower limit to
the mass of these 24$\mu m$-detected EROs as SFR$\times \tau = 50 - 170 \times 10^8
=5\times 10^9 - 2\times 10^{10} M_\odot$.
If the EROs {\it without} detectable 24$\mu m$ are indeed
massive systems with old stellar populations at $z \sim 1$ as
measured by several recent surveys (Glazebrook et al.\ 2004; Bell et
al.\ 2003), one plausible connection between the 
starburst and early-type ERO populations is that the former
may be in the process of transforming into the latter,
as initially postulated by Kormendy \&\ Sanders (1992).  
In the hierarchical clustering paradigm, it could be interpreted that
our deep 24$\mu m$ observations are capturing a massive galaxy
population in the midst of violent transformation -- possibly in the
process of assembly via mergers/starbursts at the epoch of $z \sim 1 -
2$. The accurate determination of the stellar and dynamical masses of
these starburst EROs at $z \sim 1- 2$ will be critical for the
resolution of this question.
Finally, the measurement of volume averaged mass density at $z \sim 1
- 2$ would require a better understanding of the physical source of
the integrated K-band light --- whether from dusty systems or from old stars.

Our result is consistent with what has been found with HST morphological studies
of EROs. Yan \&\ Thompson (2003) and Moustakas et al.\ (2004) have 
found that close to 50\%\ of
EROs with $K_s < 20$ have morphologies consistent with disk or later type galaxies
in the observed 8100--8500~\AA\ wavelength, and less than $40$\%\ 
show clean bulge type profiles. With the HST/ACS/NICMOS images
in the FLSV region, we will be able to investigate the morphologies of these
infrared luminous EROs, and to determine if indeed they are starbursting 
mergers at $z \sim 1$.

This work is based in part on 
observations made with the Spitzer Space Telescope, 
which is operated by the Jet Propulsion
Laboratory, California Institute of Technology under NASA
contract 1407. Support for this work was provided by NASA.
The spectroscopic data presented herein were obtained
at the W.M. Keck Observatory, which is operated as a scientific
partnership among the California Institute of Technology, the
University of California, and the National Aeronautics and Space
Administration.  The Observatory was made possible by the generous
financial support of the W.M. Keck Foundation.
We also wish to
recognize and acknowledge the very significant cultural role and
reverence that the summit of Mauna Kea has always had within the
indigenous Hawaiian community.  We are most fortunate to have the
opportunity to conduct observations from this mountain.

\clearpage
\begin{deluxetable}{lccc}
   \tablewidth{0pt}
   \tabletypesize{\scriptsize}
   \tablecaption{Summary of Optical, NIR and 24$\mu m$ Observations}
\startdata
\hline\hline \\
Filter         & $R$ & $K_s$ & 24$\mu m$ \\
FWHM ($\sec$)          & 1.0 & 1.0 & 5.5\\
Pixel Scale ($\sec$/pixel)     & 0.26 & 0.25 &  2.55 \\
$<$Exp. Time$>$ (sec)     & 1800 & 1800\tnm{a}/9000\tnm{b} & $349\tnm{a}/4268\tnm{b}$\\
3$\sigma$ Limits    & $25.5\tnm{a}/25.2\tnm{b}$ & $20.3\tnm{a}/21.0\tnm{b}$ & $90\tnm{a}/40\tnm{b}$ \\
                    &  mag  & mag  & $\mu$Jy \\
\enddata
\tablenotetext{a}{FLSV}
\tablenotetext{b}{ELAISN1}
\end{deluxetable}

\clearpage

%%%############################################################################
%%% References:

\end{document}